\documentclass[amsmath,amssymb,pra,twocolumn]{revtex4}

\usepackage{graphics}
\usepackage{epsfig}
\usepackage{amsmath}
\usepackage{color}
\usepackage{dcolumn}
\usepackage{amstext,amssymb,amsfonts,amsmath}
\usepackage{mathrsfs}
\usepackage{xspace}
\usepackage{natbib}

\newcommand{\dd}{\ensuremath{\mathrm{d}}}

\newcommand{\Aho}{\ensuremath{A_{\rm ho}}}
\newcommand{\Li}[1]{$^{#1}$Li\xspace}
\newcommand{\LiLi}{$^6$Li-$^7$Li\xspace}
\newcommand{\rel}{\rm rel.\ }
\newcommand{\cm}{\rm c.m.\ }
\newcommand{\cpl}{\rm coupl\ }
\newcommand{\reli}{\rm rel.\xspace}
\newcommand{\cmi}{\rm c.m.\xspace}

\newcommand{\cc}[1]{\ensuremath{\left(#1\right)^*}}
\newcommand{\ccb}[1]{\ensuremath{\left[#1\right]^*}}

\newcommand{\ket}[1]{\left| #1 \right>} % for Dirac bras
\newcommand{\bra}[1]{\left< #1 \right|} % for Dirac kets
\newcommand{\QMa}[3]{\left< #1 \vphantom{#2#3} \right|
 #2 \left| #3 \vphantom{#1#2} \right>} % for Dirac matrix elements
\newcommand{\mean}[1]{\left< #1 \right>} % mean value
\newcommand{\sumind}[3]{\left\{ #1,#2 \right\}\in#3} % mean value

\newcommand{\imgwidth}{0.95\linewidth}

\begin{document}

  \title{Non-perturbative theoretical description of two atoms in an optical lattice
         with time-dependent perturbations}

       \author{Philipp-Immanuel Schneider, Sergey Grishkevich, 
               and Alejandro Saenz}

       \affiliation{AG Moderne Optik, Institut f\"ur Physik,
         Humboldt-Universit\"at zu Berlin, Newtonstr. 15, 
         12489 Berlin, Germany}

       \date{\today}

  \begin{abstract}
    A theoretical approach for a non-perturbative dynamical description of two
interacting atoms in an optical lattice potential is introduced. The approach
builds upon the stationary eigenstates found by a procedure described in
Grishkevich {\it et al.} [Phys.\ Rev.\ A {\bf 84}, 062710 (2011)]. 
It allows presently to treat any time-dependent external
perturbation of the lattice potential up to quadratic order. Example
calculations of the experimentally relevant cases of an acceleration of the lattice and the turning-on of an
additional harmonic confinement are presented.
  \end{abstract}

    \maketitle

%%%%%%%%%%%%%%%%%%%%%%%%%%%%%%%%%%%%%%%%%%%%%%%%%%
 \section{Introduction}
 \label{sec:intro}
%%%%%%%%%%%%%%%%%%%%%%%%%%%%%%%%%%%%%%%%%%%%%%%%%%
%%%%%%%%%%%%%%%%%%%%%%%%%%%%%%%%%%%%%%%%%%%%%%%%%%

Triggered by the creation of the first Bose-Einstein
condensates~\cite{cold:ande95,cold:davi95}, the field of ultracold atoms has
experienced many major advancements. Nowadays it is not only possible to steer
and observe many-body effects like the Mott-insulator superfluid phase
transition~\cite{cold:grei02a,cold:bloc08,cold:bakr10} but also to manipulate single atoms
in an optical lattice (OL) or a dipole trap~\cite{cold:weit11,cold:serw11}. 

A key technology is the dynamical variation of the trapping potential that allows,
e.g., for a cooling of the system by transferring hot atoms to non-trapped
continuum states~\cite{cold:bakr11}. In a recent work by some of us it has been
proposed how to perform quantum computations in an OL just by manipulating the
depth of single lattice sites and by shaking the optical
lattice to drive transitions between different Bloch bands~\cite{cold:schn12}. 
For a full understanding of the underlying dynamical processes of any multi-band 
system \cite{cold:ande07,cold:trot08,cold:wirt10,cold:bakr11} the application of 
the usually employed single-band Hubbard model is insufficient. Here, a numerical
approach is presented, that solves the full time-dependent Schr\"odinger equation of
two interacting atoms in a single-well or multiple-well lattice, which can be
perturbed by any additional time-dependent potential up to quadratic order. 
While the types of perturbations can be easily extended, the currently implemented 
types already allow for studying many experimentally relevant situations.
For example, an acceleration of an OL or a periodic driving as realized in 
\cite{cold:ivan08,cold:sias08} results in
a linear perturbation of the lattice. The manipulation of the barrier hight between 
two lattice sites \cite{cold:ande07} or a variation of the global confinement, 
e.g. by a MOT \cite{cold:schn08}, can be simulated by adding a harmonic perturbation. 

The general problem of a precise description of interacting atoms in trapping
potentials is the existence of two very distinct length scales: that of the
short-range Born-Oppenheimer interaction (some 100 a.u.) and that of the
trapping potential (some $10\,000$ a.u.). The employed basis functions have to
cover the highly oscillating behavior in the interaction range and the slow
variation due to the trap. The use of an uncorrelated basis such as a regular 
grid or products of single-particle solutions is therefore impractical. 
A method to avoid the length scale problem
is to replace the interaction potential by a delta-like pseudo potential that
supports only a single bound state and can be adjusted to have the same $s$-wave
scattering length as the full interaction potential~\cite{cold:busc98}. In this
case the problem can be tackled, e.g., by choosing a basis of multi-band Wannier
functions of the optical lattice~\cite{cold:schn12}. However, effects of higher
partial waves or of an energy dependence of the scattering length, that can easily
become important if multiple Bloch bands are occupied~\cite{cold:schn11}, are
ignored by this approach. Here, the problem is approached by using the stationary
solutions of the lattice potential obtained by a procedure presented
in \cite{cold:gris11}. For this, the Hamiltonian is first separated into relative
(\reli) motion and center-of-mass (\cmi) motion. The different length scales are
covered by expanding the \rel and \cm wave functions in spherical harmonics and
a flexible basis of $B$ splines for the radial part \cite{cold:gris11}. In a
configuration-interaction procedure the eigenfunctions of the \rel and \cm part
of the full lattice Hamiltonian are used to determine the full eigenfunctions.
These eigenfunctions are subsequently used as a basis for the propagation
of the time-dependent wavefunction.

The paper is organized as follows. In Sec.~\ref{sec:hamiltonian} the stationary
Hamiltonian of the system is presented. In order to understand the numerical
approach, the basis functions obtained by the procedure in \cite{cold:gris11}
are shortly introduced while the interested reader should consult
\cite{cold:gris11} for a more detailed description. In Sec.~\ref{sec:TimeProp}
the time-propagation method is described. Afterwards in
Sec.~\ref{sec:CompToAnalytics} the results of the time propagation are validated
by a comparison to problems that possess an analytical solution. Finally, in
Sec.~\ref{sec:CompToLiLi} the numerical method is used to analyse a system of
\LiLi in a three-well OL. The experimentally relevant cases of an acceleration 
of the lattice, i.\,e., a linear perturbation, and of an additional harmonic 
confinement are considered.

%%%%%%%%%%%%%%%%%%%%%%%%%%%%%%%%%%%%%%%%%%%%%%%%%%
\section{Stationary Hamiltonian and its eigensolutions}
\label{sec:hamiltonian}
%%%%%%%%%%%%%%%%%%%%%%%%%%%%%%%%%%%%%%%%%%%%%%%%%%
%%%%%%%%%%%%%%%%%%%%%%%%%%%%%%%%%%%%%%%%%%%%%%%%%%

The full Hamiltonian
\begin{equation}
 \hat H(t) = \hat H_0 + \hat W(t)\\
\end{equation}
consists of a time-dependent part $\hat W(t)$ (specified below) and a stationary
part
\begin{equation}
 \hat H_0= \frac{\hat p_1^2}{2 m_1} + \frac{\hat p_2^2}{2 m_2} 
           + \hat V_{\rm lat}^{(1)}(\vec r_1\,) + \hat V_{\rm lat}^{(2)}(\vec
r_2\,)
           + \hat V_{\rm int}(\vec r_1 - \vec r_2\,)
\end{equation}
for two particles $i=1,2$ with mass $m_i$ interacting via the potential $\hat
V_{\rm int}$. In the case of ultracold atoms the isotropic interaction potential
$\hat V_{\rm int}(\vec r_1 - \vec r_2\,) =\hat V_{\rm int}(|\vec r_1 - \vec
r_2\,|)$ is described by an often only numerically given Born-Oppenheimer potential. 
The trapping potential
\begin{equation}
\label{eq:Vlat}
\hat V_{\rm lat}^{(i)} = \sum_{u = x,y,z} V_u^{(i)} \sin^2(k_u u_i)
\end{equation}
is that of an OL formed by three counter-propagating laser
beams with wave vector $k_u$ in $u$ direction ($u=x,y,z$). The lattice depth
$V_u^{(i)}$ is proportional to the laser intensity in direction $u$ and the
polarizability of particle $i$. 

\begin{figure}[htp]
 \centering
 \includegraphics[width=\imgwidth]{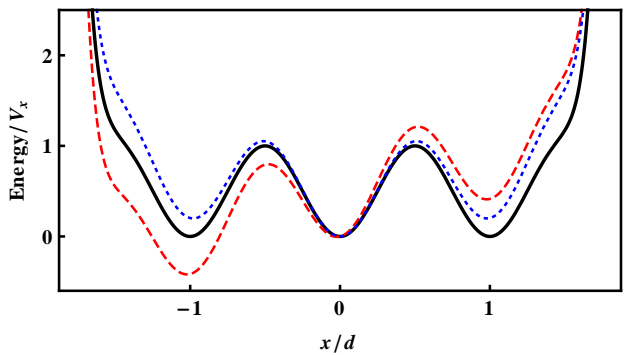}
 \caption{(color online) The 22nd-order expansion $\tilde V_{\rm
lat}(x,y=0,z=0)$ of the lattice potential
 $V_{\rm lat}(x,y=0,z=0)$ in $x$-direction (solid line).
 Lengths are given in units of the lattice spacing $d=\pi/k_x$.
 A linear perturbation as it appears, e.g., for an acceleration of the lattice in $x$ direction
 leads to an inclination of the lattice sketched by the red dashed line, while
 an additional harmonic confinement raises the left and right lattice site (blue
 dotted line).	
 }
 \label{fig:sketch}
\end{figure}

The infinite lattice potential $V_{\rm lat}$ is reduced to a potential $\tilde
V_{\rm lat}$ with finite extension by expanding $V_{\rm lat}$ to some arbitrary
order into a Taylor series in all three directions [see Fig.~\ref{fig:sketch} for the example of a
22nd order expansion of $V_x \sin^2(k_x x)$]. For practical purposes only
expansions of order $2(2n+1)$ are important, which lead to lattice potentials
$\tilde V_{\rm lat}$ with $\tilde V_{\rm lat}(\vec r\,)\rightarrow\infty$ for
$|\vec r\,|\rightarrow \infty$. For other expansions the potential is unbound 
from below leading to the appearance of non-physical states.

%%%%%%%%%%%%%%%%%%%%%%%%%%%%%%%%%%%%%%%%%%%%%%%%%%%%
%\subsection{Description of stationary eigenstates}
%%%%%%%%%%%%%%%%%%%%%%%%%%%%%%%%%%%%%%%%%%%%%%%%%%%%

The trapping potential $\hat V_{\rm lat}$ of an OL (and also $\tilde V_{\rm
lat}$) has orthorhombic symmetry, which is characterized by the point group $D_{2h}$.
By adapting the basis functions to this symmetry, the eigenfunctions and the
time-dependent wave function can be determined more efficiently. The symmetry of
the problem is discussed in depth in~\cite{cold:gris11}. Here, only the
essential points are repeated.

The symmetry operations of $D_{2h}$ are 
\begin{equation}
\label{eq:symm_operations}
 \mathcal S = \left\{ E,\; C_2(x),\; C_2(y),\; C_2(z),\; \sigma(xy),\; \sigma(xz),\;
\sigma(yz),\; i\right\}\,,
\end{equation}
where $E$ is the identity, $C_n(u)$ is the rotation about $\frac{2 \pi}{n}$
around the $u$ axis ($u=x,y,z$), $\sigma(u_1 u_2)$ the reflection on the
($u_1$, $u_2$) plane and $i$ the inversion (i.\,e.\ point reflection at the 
origin). 

Since the interaction potential $\hat V_{\rm int}$ is invariant under any operation in
$\mathcal S$ also the full unperturbed Hamiltonian $\hat H_0$ belongs to 
the $D_{2h}$ point group if the symmetry operations are performed on
both coordinates $\vec r_1$ and $\vec r_2$ simultaneously.

The group $D_{2h}$ possesses eight irreducible representations $\Gamma_\sigma$ with
\begin{equation}
\sigma \in \left\{A_g, B_{1g}, B_{2g}, B_{3g}, A_u, B_{1u}, B_{2u}, B_{3u}\right\} \,.
\end{equation}
The characters of these irreducible representations are listed in
Table~\ref{tab:CharTabD2h}.

\begin{table}[ht]
\centering
\begin{tabular}{lrrrrrrrrrr}
\hline
\hline
$D_{2h}$ & $E$ & $C_2(z)$ & $C_2(y)$ & $C_2(x)$ &   $i$  & $\sigma(xy)$ &
$\sigma(xz)$ & $\sigma(yz)$ \\    
\hline
$A_g$ & 1 & 1 & 1 & 1 & 1 & 1 & 1 & 1 \\
$B_{1g}$ & 1 & 1 & -1 & -1 & 1 & 1 & -1 & -1\\
$B_{2g}$ & 1 & -1 & 1 & -1 & 1 & -1 & 1 & -1\\
$B_{3g}$ & 1 & -1 & -1 & 1 & 1 & -1 & -1 & 1\\
$A_{u}$ & 1 & 1 & 1 & 1 & -1 & -1 & -1 & -1 \\
$B_{1u}$ & 1 & 1 & -1 & -1 & -1 & -1 & 1 & 1\\
$B_{2u}$ & 1 & -1 & 1 & -1 & -1 & 1 & -1 & 1\\
$B_{3u}$ & 1 & -1 & -1 & 1 & -1 & 1 & 1 & -1\\ 
\hline
\hline
\end{tabular}
\caption{Character table of the $D_{2h}$ point group.
}
\label{tab:CharTabD2h}
\end{table} 

In order to find the eigensolutions of $\hat H_0$ the system is split into
\rel and \cm coordinates, 
\begin{equation}
\vec{\rho}= \vec{r_1} - \vec{r_2}\,,\quad \vec{R} = \frac{m_1 \vec{r_1} + m_2
\vec{r_2}}{m_1 + m_2}. 
\end{equation}

With this separation, the Hamiltonian is written as
\begin{equation}
\hat H_0(\vec R, \vec \rho\,) = \hat H_{\cm}(\vec R) + \hat H_{\rel}(\vec \rho\,) +
\hat H_{\cpl}(\vec R,\vec \rho\,)\,,
\end{equation}
where $\hat H_{\cmi}$, $\hat H_{\reli}$, and $\hat H_{\cpl}(\vec R, \vec \rho\,)$ still have
$D_{2h}$-symmetry \cite{cold:gris11}.

The eigenfunctions of \rel and \cm are described in spherical coordinates and expanded in a
basis of $B$ splines $B_{\alpha}$ and spherical harmonics $Y_l^m$. Since the
symmetry operations of $D_{2h}$ commute with the Hamiltonian, the eigenfunctions
can be chosen such that their symmetry properties correspond to some irreducible
representation $\Gamma_\sigma$ of $D_{2h}$. In the following the \rel (\cmi)
eigenfunctions are denoted as $\phi_j^{(\sigma)}(\vec \rho\,)$
[$\Psi_j^{(\sigma)}(\vec R)$] with $j=1,2,3,\dots$

In a configuration-interaction procedure products of eigensolutions of $\hat
H_{\cmi}$ and $\hat H_{\reli}$, i.e. configurations, are used to diagonalize the full
Hamiltonian $\hat H_0$. 
Because all irreducible representations of $D_{2h}$ are one dimensional,
the direct product of two irreducible representations $\Gamma_\kappa\otimes\Gamma_\lambda$
is again an irreducible representation $\Gamma_\sigma$ that can be determined from
the product table~\ref{tab:prodD2h}.
Hence, each configuration $\Psi_i^{(\kappa)}(\vec R) \phi_j^{(\lambda)}(\vec \rho\,)$ has the 
symmetry properties of the related irreducible representation $\Gamma_\sigma = 
\Gamma_\kappa\otimes\Gamma_\lambda$.
 The full solutions of a given symmetry $\sigma$ has the form of a superposition 
\begin{equation}
\label{eq:CISol}
 \Phi_\sigma(\vec R,\vec \rho\,) = \sum_{\sumind{\kappa}{\lambda}{\sigma}} \sum_{i j} \mathcal C_{i j}^{(\kappa,
\lambda)} \Psi_i^{(\kappa)}(\vec R) \phi_j^{(\lambda)}(\vec \rho\,)\,,
\end{equation}
where $\sumind{\kappa}{\lambda}{\sigma}$ should indicate that the summation is performed over
irreducible representations that fulfill $\Gamma_\kappa\otimes\Gamma_\lambda=\Gamma_\sigma$. 

When considering identical bosonic
(fermionic) particles the \rel wavefunction has to be symmetric (antisymmetric)
under inversion, i.\,e.\ only basis functions of \rel motion with $\lambda \in
\{A_g, B_{1g}, B_{2g}, B_{3g}\}$ ($\lambda \in \{A_u, B_{1u}, B_{2u}, B_{3u}\}$) are
used to form configurations.

The wavefunctions, i.\,e.\ the coefficients $ C_{i j}^{(\kappa, \lambda)}$ in 
Eq.~\eqref{eq:CISol}, are finally determined by solving the eigenvalue problem 
\begin{equation}
\label{eq:EVP_CI}
\hat H_0 \ket{\Phi_i^{(\sigma)}}
= E_i^{(\sigma)} \ket{\Phi_i^{(\sigma)}}
\end{equation}
of $\hat H_0$ in the configuration basis.

\begin{table}[ht]
\centering
\begin{tabular}{lrrrrrrrrrr}
\hline
\hline
$\otimes$ & $A_g$ & $B_{1g}$ & $B_{2g}$ & $B_{3g}$ &  $A_u$  & $B_{1u}$ & $B_{2u}$ &
$B_{3u}$ \\    
\hline
$A_g$ & $A_g$ & $B_{1g}$ & $B_{2g}$ & $B_{3g}$ &  $A_u$  & $B_{1u}$ & $B_{2u}$ &
$B_{3u}$ \\
$B_{1g}$ & $B_{1g}$ & $A_g$ & $B_{3g}$ & $B_{2g}$ & $B_{1u}$ & $A_u$ & $B_{3u}$
& $B_{2u}$\\
$B_{2g}$ & $B_{2g}$ & $B_{3g}$ & $A_g$ & $B_{1g}$ & $B_{2u}$ & $B_{3u}$ & $A_u$
& $B_{1u}$\\
$B_{3g}$ & $B_{3g}$ & $B_{2g}$ & $B_{1g}$ & $A_g$ & $B_{3u}$ & $B_{2u}$ &
$B_{1u}$ & $A_u$ \\
$A_{u}$ & $A_{u}$ & $B_{1u}$ & $B_{2u}$ & $B_{3u}$ & $A_g$ & $B_{1g}$ & $B_{2g}$
& $B_{3g}$\\
$B_{1u}$ & $B_{1u}$ & $A_u$ & $B_{3u}$ & $B_{2u}$ & $B_{1g}$ & $A_g$ & $B_{3g}$
& $B_{2g}$\\
$B_{2u}$ & $B_{2u}$ & $B_{3u}$ & $A_u$ & $B_{1u}$ & $B_{2g}$ & $B_{3g}$ & $A_g$
& $B_{1g}$\\
$B_{3u}$ & $B_{3u}$ & $B_{2u}$ & $B_{1u}$ & $A_u$ & $B_{3g}$ & $B_{2g}$ &
$B_{1g}$ & $A_g$\\ 
\hline
\hline
\end{tabular}
\caption{Product table of irreducible representations of the $D_{2h}$ point group.}
\label{tab:prodD2h}
\end{table} 

%%%%%%%%%%%%%%%%%%%%%%%%%%%%%%%%%%%%%%%%%%%%%%%%%%
\section{Time-dependent evolution}
\label{sec:TimeProp}
%%%%%%%%%%%%%%%%%%%%%%%%%%%%%%%%%%%%%%%%%%%%%%%%%%
%%%%%%%%%%%%%%%%%%%%%%%%%%%%%%%%%%%%%%%%%%%%%%%%%%

The Schr\"odinger equation of the time-dependent evolution 
\begin{equation}
\label{eq:TimeDepSchr}
\begin{split}
 \left(\hat H_0 + \hat W(t)\right)\ket{\Psi(t)} &= i \hbar
\frac{\partial}{\partial t} \ket{\Psi(t)}\\
 \text{with} \quad \ket{\Psi(t=0)}&=\ket{\Psi_0}
\end{split}
\end{equation}
is solved in the basis $\{\Phi_i^{(\sigma)}\}$ of eigenfunctions of $\hat H_0$ of Eq.~\eqref{eq:EVP_CI},
\begin{equation}
\label{eq:BasTimeDepWF}
 \ket{\Psi(t)} = \sum_{\sigma i} \mathcal B_{\sigma i}(t)
\ket{\Phi_i^{(\sigma)}}\,.
\end{equation}

Plugging Eq.~\eqref{eq:BasTimeDepWF} into Eq.~\eqref{eq:TimeDepSchr} 
and multiplying from the left by $\bra{\Phi_j^{(\kappa)}}$ leads to the equation
\begin{equation}
i \hbar \frac{\partial \mathcal B_{\kappa j}(t)}{\partial t} = E_j^{(\kappa)}
\mathcal B_{\kappa j}(t) + \sum_{\sigma i} \mathcal B_{\sigma i}(t)
\bra{\Phi_j^{(\kappa)}} \hat W(t) \ket{\Phi_i^{(\sigma)}}
\end{equation}
for the evolution of the time-dependent coefficients  $\mathcal B_{\kappa
j}(t)$, which is governed by the matrix
elements $\mathcal P_{i j}^{(\kappa, \sigma)} = \bra{\Phi_j^{(\kappa)}} \hat W(t)
\ket{\Phi_i^{(\sigma)}}$ of the perturbation.
Considering two eigenstates
\begin{equation}
\begin{split}
  \ket{\Phi_m^{(\tau)}} =& \sum_{\sumind{\kappa}{\lambda}{\tau}} \sum_{i j} \mathcal C_{i j}^{(\kappa,
\lambda)} \ket{\Psi_i^{(\kappa)}} \ket{\phi_j^{(\lambda)}}\\
  \ket{\Phi_n^{(\sigma)}} =& \sum_{\sumind{\mu}{\nu}{\sigma}} \sum_{k l} \mathcal{C'}_{k l}^{(\mu,
\nu)} \ket{\Psi_k^{(\mu)}} \ket{\phi_l^{(\nu)}}\\
\end{split} 
\end{equation}
as specified in Eq.~\eqref{eq:CISol}, the matrix elements of a perturbation 
%of the form $W(t) = f(t) \hat W$ 
are
\begin{equation}
\label{eq:general_P}
\begin{split}
 \mathcal P_{m n}^{(\tau, \sigma)} =& \bra{\Phi_m^{(\tau)}} \hat W(t)
\ket{\Phi_n^{(\sigma)}}\\
=& % f(t)
\sum_{\sumind{\kappa}{\lambda}{\tau}} \sum_{i j}
\sum_{\sumind{\mu}{\nu}{\sigma}} \sum_{k l}  
\cc{\mathcal {C}_{i j}^{(\kappa, \lambda)}}
\mathcal{C'}_{k l}^{(\mu,\nu)}\\
&\times\bra{\phi_j^{(\lambda)}}\bra{\Psi_i^{(\kappa)}} \hat W(t)
\ket{\Psi_k^{(\mu)}}\ket{\phi_l^{(\nu)}}\,.
\end{split}
\end{equation}

In general, the perturbation $\hat W(t)$ can be expanded
in a time-dependent Taylor series of its spacial coordinates 
\[
\hat W(t) = \sum_{n m}\sum_{u=x,y,z}\sum_{u'=x,y,z} f_{n m}^{(u,u')}(t)\, \hat R_u^n\, \hat \rho_{u'}^m\,,
\]
where $\rho_u$ ($R_u$) is the component of the \rel (\cmi) motion
in $u$ direction ($u=x,y,z$).

At the present stage perturbations in $x$ direction of the general form
\begin{equation}
\label{eq:perturbation}
\begin{split}
 \hat W(t) =& f_{01}(t)\, \hat \rho_x + f_{10}(t)\, \hat R_x + f_{11}(t)\, \hat \rho_x \hat R_x
\\
            &+ f_{02}(t)\, \hat \rho_x^2 + f_{20}(t)\, \hat R_x^2  
\end{split}
\end{equation}
are implemented. In principle, the method can be easily extended to allow for 
perturbations in other directions and of higher orders.

In order to illustrate how the perturbation matrix is computed, the case of a
linear perturbation $\hat W = f_{10}(t) \hat R_x$ is discussed in more detail. This perturbation
does not couple the orthonormal \rel basis functions $\ket{\phi_j^{(\lambda)}}$. Thus, the summations in
Eq.~\eqref{eq:general_P} reduce to
\begin{equation}
\begin{split}
 \mathcal P_{m n}^{(\tau, \sigma)} =  &f_{10}(t)
\sum_{\sumind{\kappa}{\lambda}{\tau}} \sum_{i j} 
\sum_{\sumind{\mu}{\lambda}{\sigma}} \sum_{k}  
\cc{\mathcal {C}_{i j}^{(\kappa, \lambda)}}
\mathcal{C'}_{k j}^{(\mu, \lambda)}\\
&\bra{\Psi_i^{(\kappa)}} \hat R_x \ket{\Psi_k^{(\mu)}}\,.
\end{split}
\end{equation}

In the following the term $\bra{\Psi_i^{(\kappa)}} \hat R_x
\ket{\Psi_k^{(\mu)}}$
is considered for the exemplary case of $\kappa=A_g$. In this case the wave function
$\Psi_i^{(\kappa)}(\vec R\,)$ is totally symmetric (see Table~\ref{tab:CharTabD2h}). 
Hence, $\Psi_k^{(\mu)}(\vec R\,)$ needs to be anti-symmetric in $x$ direction and symmetric otherwise, 
which is fulfilled solely for $\mu = B_{3u}$. In all other cases the integral vanishes.
The according \cm basis functions are represented as
\begin{equation}
\begin{split}
 \Psi_{i}^{(A_g)}(R,\Theta,\Phi) &= \sum_{\alpha = 1}^{N_\alpha} \sum_{l=0,\{2\}}^{N_l} \sum_{m=0,\{2\}}^l
 c_{i, \alpha l m}^{(A_g)} \frac{B_\alpha(R)}{R} \mathscr Y_{l m}^+\\
 \Psi_{k}^{(B_{3u})}(R,\Theta,\Phi) &= \sum_{\alpha = 1}^{N_\alpha} \sum_{l=1,\{2\}}^{N_l} \sum_{m=1,\{2\}}^l
 c_{k, \alpha l m}^{(B_{3u})} \frac{B_\alpha(R)}{R} \mathscr Y_{l
m}^-\, ,
\end{split}
\end{equation}
where $B_\alpha$ are $B$ splines, $\mathscr Y_{l m}^\pm = Y_l^m(\Theta,\Phi) \pm Y_l^{-m}(\Theta,\Phi)$ are a sum of spherical harmonics for $m\neq 0$,
and $\mathscr Y_{l 0}^\pm = Y_l^0(\Theta,\Phi)$ (see \cite{cold:gris11} for details). 
The numbers in curly brackets below the sums indicate the summation step.
$N_\alpha$ and $N_l$ are variable values of the number of $B$ splines and the maximal angular momentum, respectively. 
With $R_x = R \sin\Theta\cos\Phi$ one finds
\begin{equation}
\label{eq:FullIntegral}
\begin{split}
\bra{\Psi_i^{(\kappa)}}& \hat R_x \ket{\Psi_k^{(\mu)}} \\
=&  
\sum_{l=0,\{2\}} \sum_{m=0,\{2\}}^l
\sum_{l'=1,\{2\}} \sum_{m'=1,\{2\}}^{l'} \sum_{\alpha \alpha'} \cc{c_{i,
\alpha l m}^{(A_g)}} c_{k, \alpha l m}^{(B_{3u})} \\
&\times \int \dd R\; B_\alpha(R) R B_{\alpha'}(R) \\
&\times \int_0^{\pi} \sin \Theta\, \dd \Theta\; \int_0^{2\pi} \dd \Phi\; 
\cc{\mathscr Y_{l m}^+} 
\sin\Theta \cos\Phi \mathscr Y_{l m}^-\,.
\end{split}
\end{equation}
Using the identities $\cc{Y_l^m} = (-1)^m
Y_l^{-m}$,
%\begin{equation}
$\sin \Theta \cos\Phi =
\sqrt{\frac{2\pi}{3}}\left[Y_1^{-1}(\Theta,\Phi)-Y_1^{1}(\Theta,\Phi)\right]$,
%\end{equation}
and
\begin{equation}
\begin{split}
 \int_0^{\pi} \sin \Theta\, \dd \Theta\; \int_0^{2\pi} \dd \Phi\;
Y_{l_1}^{m_1}(\Theta,\Phi) Y_{l_2}^{m_2}(\Theta,\Phi) Y_{l_3}^{m_3}(\Theta,\Phi)
= \\
\sqrt{\frac{(2l_1+1)(2l_3+1)(2l_3+1)}{4\pi}}
\begin{pmatrix}
 l_1 & l_2 & l_3\\
 0   & 0   & 0
\end{pmatrix}
\begin{pmatrix}
 l_1 & l_2 & l_3\\
 m_1 & m_2 & m_3
\end{pmatrix}\, , 
\end{split}
\end{equation}
the integral over the angles in Eq.~\eqref{eq:FullIntegral} can be efficiently computed in terms of
Wigner 3j-symbols $\begin{pmatrix}
 l_1 & l_2 & l_3\\
 m_1 & m_2 & m_3
\end{pmatrix}$. 
%is the Wigner 3j-symbol. With $\cc{Y_l^m} = (-1)^m
%Y_l^{-m}$ and 
%\begin{equation}
%\sin \Theta \cos\Phi =
%\sqrt{\frac{2\pi}{3}}\left[Y_1^{-1}(\Theta,\Phi)-Y_1^{1}(\Theta,\Phi)\right] 
%\end{equation}
%this yields
%%
%\begin{equation}
%\begin{split}
% \int_0^{\pi} \sin \Theta\, &\dd \Theta\; \int_0^{2\pi} \dd \Phi\;
%\ccb{Y_{l_1}^{m_1}(\Theta,\Phi)} \sin \Theta \cos\Phi
%Y_{l_2}^{m_2}(\Theta,\Phi)\\
%=&
%(-1)^{m_1} \sqrt{\frac{(2l_1+1)(2l_2+1)}{2}}
%\begin{pmatrix}
% l_1 & l_2 & 1\\
% 0   & 0   & 0
%\end{pmatrix}\\
%&\times\left(
%\begin{pmatrix}
% l_1 & l_2 & 1\\
% -m_1 & m_2 & -1
%\end{pmatrix} 
%-
%\begin{pmatrix}
% l_1 & l_2 & 1\\
% -m_1 & m_2 & 1
%\end{pmatrix} 
%\right)\, ,
%\end{split}
%\end{equation}
%which can be efficiently computed. 
The other types of perturbations in Eq.~\eqref{eq:perturbation} are treated in an analogous way.

Since the system is six dimensional the analysis in terms of the full 
time-dependent wavefunction is nontrivial. However, equipped with the matrix elements
of all perturbations, $\bra{\Phi_i^{(\kappa)}} \hat R_x \ket{\Phi_k^{(\mu)}}$,
$\bra{\Phi_i^{(\kappa)}} \hat R_x^2 \ket{  \Phi_k^{(\mu)}}$,
$\bra{\Phi_i^{(\kappa)}} \hat \rho_x \ket{ \Phi_k^{(\mu)}}$,
$\bra{\Phi_i^{(\kappa)}} \hat \rho_x^2\ket{\Phi_k^{(\mu)}}$,
and
$\bra{\Phi_i^{(\kappa)}} \hat R_x \hat \rho_x \ket{\Phi_k^{(\mu)}}$ one can easily 
determine the expectation values of some of the most important observables.
For example, the squared mean particle distance in $x$ direction is given as
\[
\begin{split}
\mean{\rho_x^2} =& \QMa{\Psi(t)}{ \hat \rho_x^2 }{\Psi(t)}\\
=&\sum_{\sigma i}\sum_{\kappa j} \ccb{\mathcal B_{\sigma i}(t)} \mathcal B_{\kappa j}(t)
\bra{\Phi_i^{(\sigma)}} \hat \rho_x^2\ket{\Phi_j^{(\kappa)}}\,.
\end{split}
\]
Likewise, one can determine the mean particle position or the uncertainty of the position
in $x$ direction.

%%%%%%%%%%%%%%%%%%%%%%%%%%%%%%%%%%%%%%%%%%%%%%%%%%
 \section{Comparison with analytical results}
 \label{sec:CompToAnalytics}
%%%%%%%%%%%%%%%%%%%%%%%%%%%%%%%%%%%%%%%%%%%%%%%%%%
%%%%%%%%%%%%%%%%%%%%%%%%%%%%%%%%%%%%%%%%%%%%%%%%%%

 In order to validate the numerical procedure a comparison with analytical
results is necessary, which are available for the harmonic approximation of the
OL potential. In the case of two identical particles of mass $m$ in a harmonic trap 
the system decouples into \rel and \cm motion
with Hamiltonian
\begin{equation}
 \hat H_0 = \frac{\hat P^2}{2 M} + \frac{1}{2} M \omega^2 R^2 + \frac{\hat
p^2}{2 \mu} + \frac{1}{2} m \omega^2 \rho^2 + V_{\rm int}(\rho)\,.
\end{equation}
Here, $M=2m$, $\mu=m/2$, $\hat P$ is the momentum of \cm and
$\hat p$ the momentum of \rel motion. In the following, a linear perturbation
$\hat W(t) = f(t) \hat R_x$ and a quadratic perturbation $\hat W(t) = f(t) \hat R_x^2$, i.\,e.\ a time-dependent 
acceleration and a variation of the trapping frequency, are considered. 
Since the \cm part of $H_0$ decouples into $x,y$, and $z$ direction, only
the \cm harmonic oscillator in $x$ direction with Hamiltonian 
\begin{equation}
 \hat H_{\rm ho} = \frac{\hat P_x^2}{2 M} + \frac{1}{2} M \omega^2 \hat R_x^2
  =\hbar\omega\left(\Aho^2 \frac{\hat P^2}{2 \hbar^2} + \frac12
\frac{\hat R_x^2}{\Aho^2}\right)
\end{equation}
is affected by the perturbations, where $\Aho = \sqrt{\hbar/(M \omega)}$ is the
harmonic oscillator length. 
The comparisons are performed for expectation values of the position
\begin{equation}
\label{eq:X_0}
 \bar X (t) = \mean{\hat R_x} = \QMa{\Psi(t)}{\hat R_x}{\Psi(t)}
\end{equation}
and the mean deviation from $\bar X$
\begin{equation}
\label{eq:sigma}
 \sigma(t) = \sqrt{\mean{\hat R_x^2}-\mean{\hat R_x}^2} \,.
\end{equation}

%%%%%%%%%%%%%%%%%%%%%%%%%%%%%%%%%%%%%%%%%%%%%%%%%%
\subsection{Periodic driving}
%%%%%%%%%%%%%%%%%%%%%%%%%%%%%%%%%%%%%%%%%%%%%%%%%%

For the case of a periodically driven harmonic oscillator 
with driving strength $C_{\rm shake}$ and frequency $\omega_0$,
\begin{equation}
\hat W_1(t) = \hbar \omega\, C_{\rm shake}\,\cos\left(\omega_0
t\right)\,\frac{\hat R_x}{\Aho}\,,
\end{equation}
there exists an analytic solution \cite{cold:band08},
\begin{equation}
\psi_n(X,t)=e^{i \varphi(x,t)}\phi_n(X-\xi(t))\,,
\end{equation}
where $\varphi(x,t)$ is a phase, which vanishes for $t=0$,
$\phi_n$ is the $n$th harmonic oscillator eigenstate of $\hat H_{\rm ho}$, and
\begin{equation}
\xi(t)=\frac{\Aho C_{\rm shake} }{1-\omega_0^2/\omega^2}\cos(\omega_0 t).
\end{equation}
In order to conform with the initial condition
\begin{equation}
\psi_n(X,0)=\phi_n(X-\xi(0))
\end{equation}
the trap is shifted at $t=0$ to $\xi(0)$ by instantly adding a constant linear
perturbation 
\begin{equation}
\hat W_2=-\hbar\omega\, C_{\rm shake}\,
\frac{1}{1-\omega_0^2/\omega^2}\, \frac{\hat R_x}{\Aho} \,.
\end{equation}

From the analytic solution one obtains straightforwardly
\begin{equation}
\label{eq:X_0_analytic}
\bar X(t) = - \Aho C_{\rm shake} \left[1 - \cos(\omega_0 t)\right], \quad \sigma(t)
=  \frac{\Aho}{\sqrt{2}}.
\end{equation}

\begin{figure}[htp]
 \centering
 \includegraphics[width=\imgwidth]{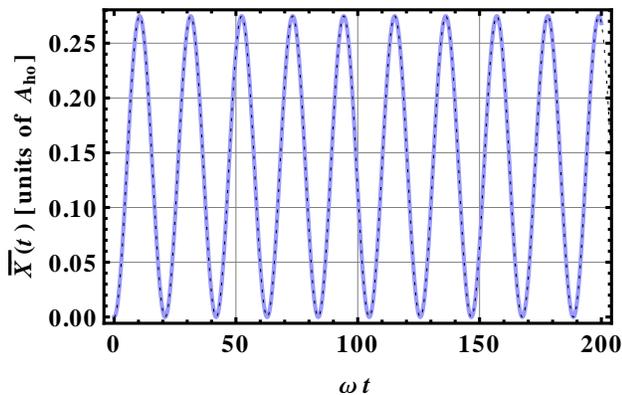}
 \caption{(color online) Comparison of analytical (blue solid) and numerical
(black dashed) results for $\bar X(t)$ [see Eqs.~\eqref{eq:X_0} and 
\eqref{eq:X_0_analytic}] for $C_{\rm shake} = 0.5$. The difference of the
results is below $10^{-10}$ and therefore invisible. The width of the wave
function $\sigma(t) =  \Aho/\sqrt{2}$ is numerically reproduced with
the same level of accuracy.
  }
 \label{fig:X0_CM}
\end{figure}

In Fig.~\ref{fig:X0_CM} a comparison of a numerical calculation  of $\bar X(t)$ 
to the result in Eq.~\eqref{eq:X_0_analytic} shows very
good agreement with deviations on the order of $10^{-10}$. 
A similar accuracy is obtained for the value of $\sigma(t)$.
The deviations are due to the finiteness of the basis which, in the shown calculation,
only includes basis functions with an eigenenergy below the chosen cutoff of $20\hbar\omega$. The energy 
cutoff can be adapted to reach higher accuracies, if needed.

%%%%%%%%%%%%%%%%%%%%%%%%%%%%%%%%%%%%%%%%%%%%%%%%%%
\subsection{Adiabatic deepening}
%%%%%%%%%%%%%%%%%%%%%%%%%%%%%%%%%%%%%%%%%%%%%%%%%%

The mean width of the wavefunction $\sigma$ for an harmonic oscillator with oscillator length
$\Aho$ is given as $\Aho/\sqrt{2}$ [see Eq.~\eqref{eq:X_0_analytic}]. Considering a time dependent
perturbation $\hat W(t)= C_{\rm harm} \hbar \omega \hat R_x^2/\Aho^2 \omega t $, 
the full potential is given as $\frac{1}{2} \hbar \omega 
\frac{R_x^2} {\Aho^2}(1+2 C_{\rm harm} \omega t)$. If the
perturbation happens sufficiently slowly, the wave function will always remains in
an eigenstate of a harmonic oscillator with a trap length
\begin{equation}
\label{eq:A_t}
 \Aho(t) = \Aho(t = 0)/\sqrt{1+2 C_{\rm harm} \omega t} \,.
\end{equation}
Thus, assuming perfect adiabaticity, the width of the wave function behaves like
\begin{equation}
\label{eq:sigma_analytic}
 \sigma(t) = \Aho/\sqrt{2(1+2 C_{\rm harm} \omega t)}\,.
\end{equation}

\begin{figure}[htp]
 \centering
 \includegraphics[width=\imgwidth]{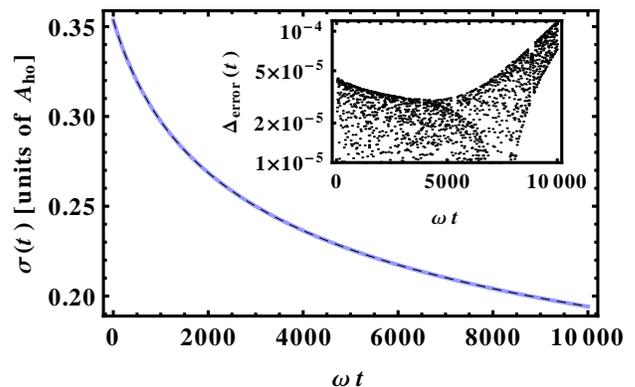}
 \caption{(color online) Comparison of analytical (blue solid) and numerical
(black dashed) results for $\sigma(t)$ [see Eqs.~\eqref{eq:sigma} and
\eqref{eq:sigma_analytic}] for $C_{\rm harm} = 0.002$. The error $\Delta_{\rm
error} = |\sigma - \sigma_{\rm num}|$ is shown in the inset. The relatively
large error in comparison to the results shown in Fig.~\ref{fig:X0_CM} is due to
nonadiabatic effects. These effects get smaller for larger $t$
since the change of $\Aho(t)$ is reduced [see Eq.~\eqref{eq:A_t}]. For
$\omega t > 5000$,  however, the incompleteness of the basis used for the
numerical calculations (only states with energies below $E=20\,\hbar\omega$ are
included) leads finally to an increase of $\Delta_{\rm error}$.
  }
 \label{fig:Sigma_CM}
\end{figure}

In Fig.~\ref{fig:Sigma_CM} a comparison to the numerical calculations shows good
agreement to this result with an error of about $5\times 10^{-5}$ for $C_{\rm harm}=0.002$, which is due
to nonadiabatic effects. For example, reducing the speed of the perturbation by
setting $C_{\rm harm}=0.001$ reduced the error to about $2\times 10^{-5}$.

%%%%%%%%%%%%%%%%%%%%%%%%%%%%%%%%%%%%%%%%%%%%%%%%%%
 \section{Example calculations for \LiLi}
\label{sec:CompToLiLi}
%%%%%%%%%%%%%%%%%%%%%%%%%%%%%%%%%%%%%%%%%%%%%%%%%%
%%%%%%%%%%%%%%%%%%%%%%%%%%%%%%%%%%%%%%%%%%%%%%%%%%

In the following a system of two distinguishable atoms, $^6$Li and $^7$Li is considered. The
interaction potential $V_{\rm int}$ is given by the Born-Oppenheimer potential
for scattering of spin-polarized lithium. 
The atoms are confined in a three-site lattice potential $\tilde V_{\rm
lat}$, which is realized by a 22nd order expansion of $V_{\rm lat}$ in
Eq.~\eqref{eq:Vlat} in $x$ direction (see Fig.~\ref{fig:sketch}) and a harmonic approximation 
in $y$ and $z$ direction. 
The chosen wave vectors $k_x=k_y=k_z=2\pi/(1000\,{\rm nm})$ lead to a lattice spacing of
$d=500\,{\rm nm}=9450\,$a.u.
A lattice depth in $x$ direction of $V_x=1.36 \hbar\omega_1$, where $\omega_1$ 
is the frequency of the harmonic approximation of the lattice for atom 1 
($^6$Li), results in the relatively small hopping energies  
$J_1 = 2.1 \times 10^{-4} \hbar\omega_1$ of atom 1 and 
$J_2 = 1.5 \times 10^{-4} \hbar\omega_1$ of atom 2 in the corresponding Hubbard model 
for the infinite lattice.
Hence, even for the relatively small $s$-wave scattering length of $41\,$a.u.\ 
of \LiLi a correlated Mott-like state is formed, i.\,e., the atoms do not occupy 
the same lattice site in the ground state \footnote{The occupation of
the deeply bound molecular states is neglected during the calculation.}. 
Since no unit filling of the lattice is considered, the atoms are nevertheless
mobile in $x$ direction. 
This enables the observation of a correlated motion of the distinguishable atoms.
The lattice depths in $y$ and $z$ direction are given as
$V_y=V_z \approx 8 V_x$ such that for low-lying states motion in these
directions is frozen out.

Despite the reduction to only three lattice sites, the considered system exhibits 
the basic mechanisms of hopping and onsite interaction of atoms in an OL.
Similar systems of only a few lattice sites appear also experimentally in 
superlattices \cite{cold:ande07}.

%%%%%%%%%%%%%%%%%%%%%%%%%%%%%%%%%%%%%%%%%%%%%%%%%%
 \subsection{Linear perturbation}
%%%%%%%%%%%%%%%%%%%%%%%%%%%%%%%%%%%%%%%%%%%%%%%%%%

\begin{figure}[htp]
 \centering
(a)\hfill\phantom{x}
\includegraphics[width=\imgwidth]{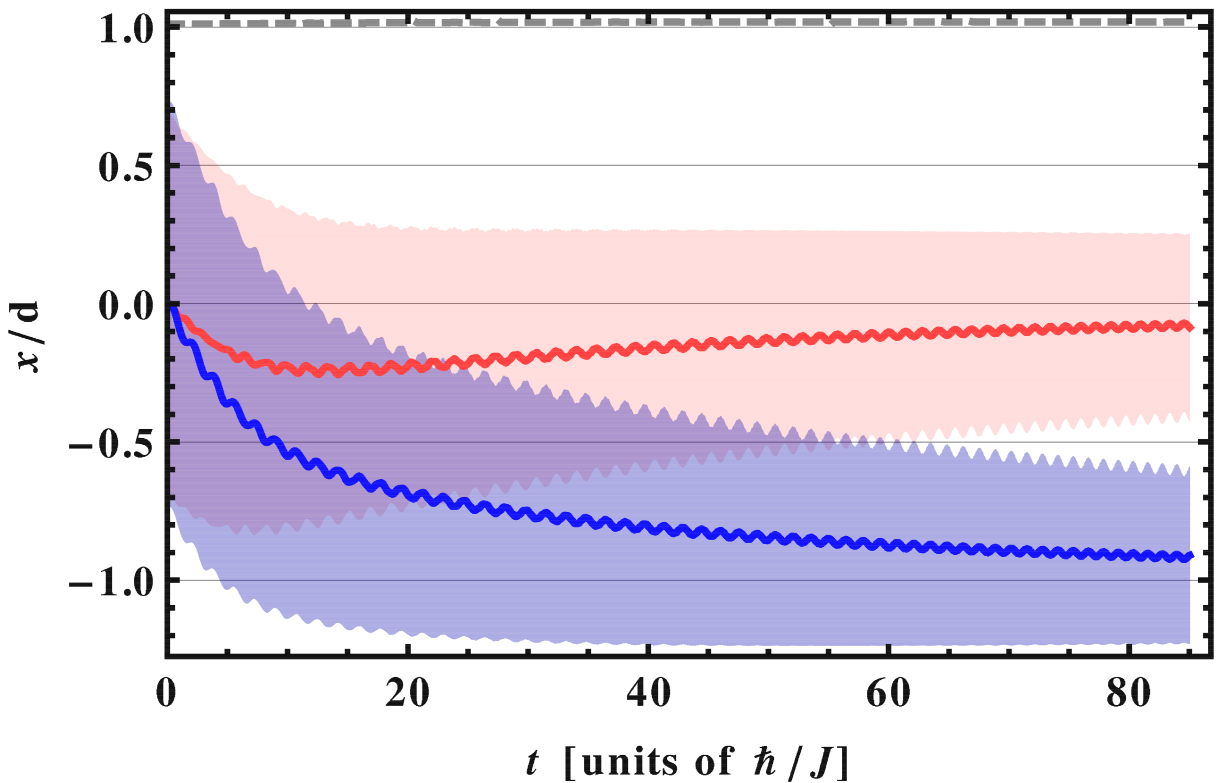}
(b)\hfill\phantom{x}\\
\includegraphics[width=0.48\linewidth]{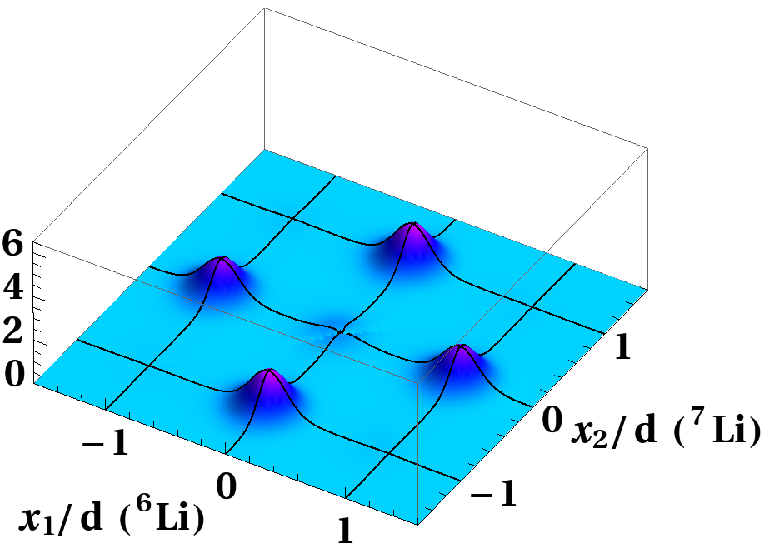}\hfill
\includegraphics[width=0.48\linewidth]{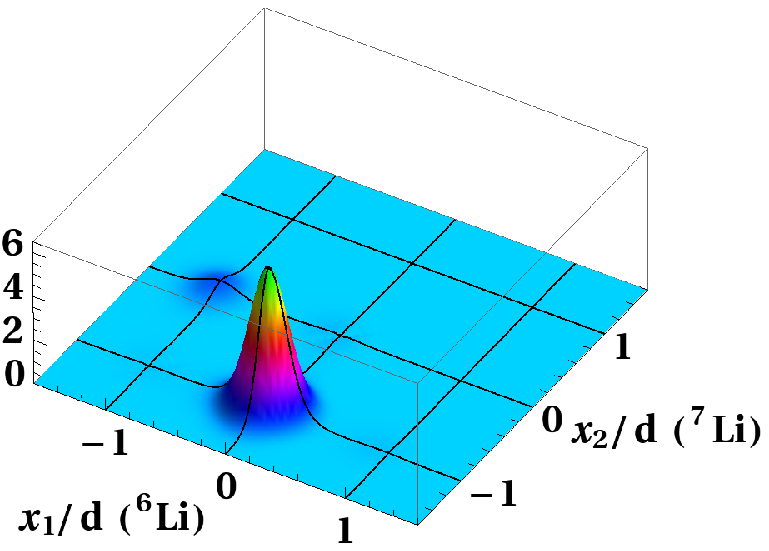}\\
(c)\hfill\phantom{x}
\includegraphics[width=\imgwidth]{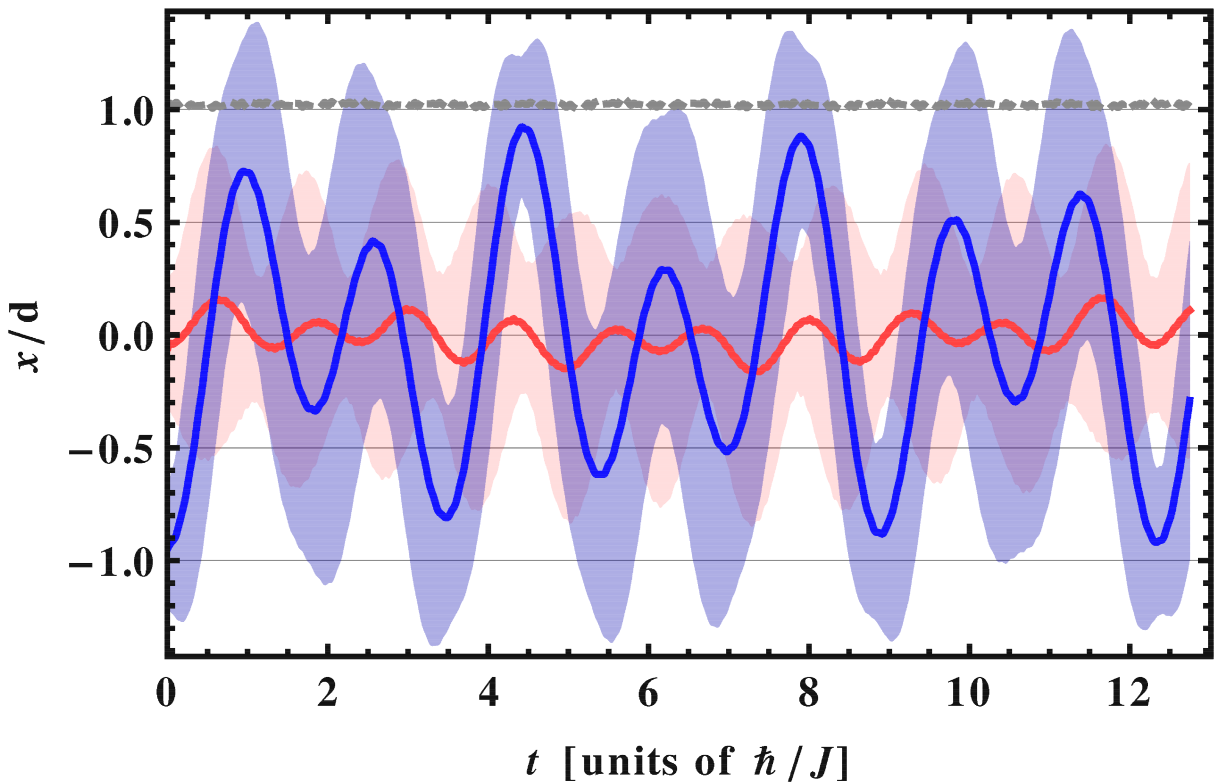}
 \caption{(color online) Mean particle position $\bar x_i = \mean{x_i}$
 of \Li{6} (thick lighter red line) and of \Li{7} (thick darker blue line) and
 mean distance $\sqrt{\mean{\rho_x^2}}$ (grey dashed line).
 The corresponding lighter red, and darker blue shading illustrates the
uncertainty 
 of the position $\bar x_i \pm \sqrt{\mean{(x_i-\bar x_i)^2}}$ of  \Li{6} and  
 \Li{7}, respectively. Time is given in units of the hopping time $\hbar/J_1$ of
\Li{6}.
 (a) Time dependent behavior for a linear inclination with a final perturbation 
     $\hat W = 66 J_1 \hat R_x/d = 0.09\, \hbar\omega_1 \hat R_x/d$.
 (b) Probability density $|\Psi(x_1,x_2)|^2$ for $y_1=y_2=z_1=z_2=0$ of the initial
     state (left) and the final state (right). Initially there is an almost
     equal probability of finding $^6$Li in the central well and $^7$Li in the outer wells
     and vice versa.
     After the linear inclination $^7$Li is predominantly situated in the left well
     and $^6$Li in the central well. The situation with exchanged $^6$Li and $^7$Li
     the wavefunction has a small but non-vanishing probability.
 (c) Free evolution of the system with the initial state being the final
     state of the process in (a). }
 \label{fig:incline}
\end{figure}

First, the system is adiabatically inclined by a perturbation of the type $\hat
W(t) = A t \hat R_x$ [see Fig.~\ref{fig:incline} (a)]. Experimentally this could,
e.g., be realized by slowly increasing the acceleration of the lattice in
$x$ direction. The system starts in the ground state where
the atoms spread symmetrically over the lattice. As a consequence, the mean atom 
position is exactly in the middle of the triple-well potential, i.\,e.\ at $x/d=0$. 
Due to their repulsion the atoms never occupy
the same lattice site. In this case their mean distance $\sqrt{\mean{\rho_x^2}}$
is approximately $d$. The corresponding probability density along the $x$ axis
is shown in the left graph of Fig.~\ref{fig:incline} (b).

Upon inclining, the system stays in the state of minimal energy, i.\,e.\ the 
heavier \Li{7} atom slowly moves into the lower left lattice site 
(i.\,e.\ $\bar x_2 = \mean{x_2}$ approaches $-d$) while the
lighter \Li{6} atom moves to the central site (i.\,e.\ $\bar x_1=\mean{\hat x_1}$
approaches zero), where it avoids an energy gain due to the interatomic repulsion.
With much smaller probability the same process with exchanged $^6$Li and $^7$Li appears 
[see right graph of Fig.~\ref{fig:incline} (b)]. 
During the process the mean distance is unchanged while the
uncertainty of the position $\sqrt{\mean{(x_i-\bar x_i)^2}}$ of atom $i$ ($i=1,2$)
decreases [see Fig.~\ref{fig:incline} (a) and (b)].
Stopping at a final inclination that results in an energy difference of 
$0.09 \hbar\omega$ between neighboring wells, the atoms are well separated. 
For a further inclination both $^6$Li and $^7$Li would move to the left well.

Starting from a system of separated atoms, one can induce a collision process.
To this end the linear perturbation, i.\,e.\ the acceleration, is suddenly switched off. 
As shown Fig.~\ref{fig:incline} (c) in this case the heavier
atom tunnels back and forth between the left and the right well.
Due to the small initial population of the state where $^6$Li is in the left well
and $^7$Li in the central well, also $^6$Li tunnels back and forth and 
$\bar x_1$ oscillates slightly around zero. 
Owing to the mass difference both tunneling processes of $^6$Li and $^7$Li happen 
with different frequencies.
Due to the repulsion during the tunneling process 
the atoms do still not occupy the same lattice site which is obvious from the unchanged
particle distance.

\begin{figure}[htp]
 \centering
 \includegraphics[width=\imgwidth]{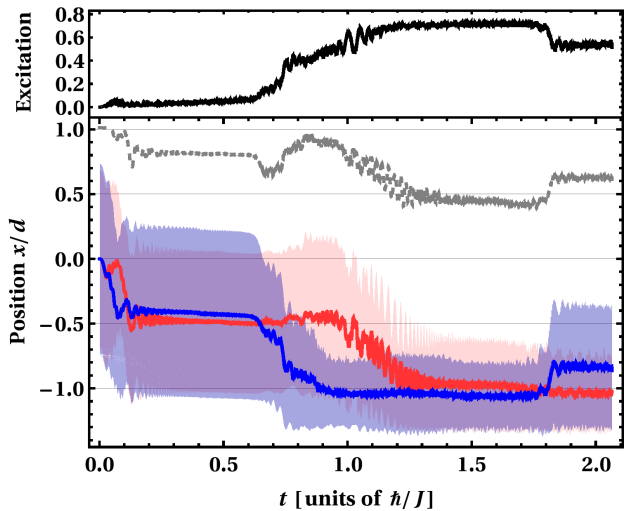} 
 \caption{(color online) Time dependent behavior for a linear inclination with a
final perturbation 
 $\hat W = 4400\, J_1 \hat R_x/d =0.6\, \hbar\omega_1 \hat R_x/d$.
 Top: Total occupation probability of states above the first Bloch band. 
 Bottom: Legend as in Fig.~\ref{fig:incline}.
 }
 \label{fig:fastincline}
\end{figure}

While a weak adiabatic inclination can be easily described also within the
standard Hubbard model, a fast inclination couples states of different Bloch
bands \cite{cold:schn12}. In Fig.~\ref{fig:fastincline} the behavior for a
stronger and faster inclination than the one in Fig.~\ref{fig:incline} (a) is
presented. In this case the behavior is harder to predict. For example, it is unclear
whether either first the heavier atom or the lighter atom moves to the left
lattice site. Although one could expect that the lighter atom with its larger
tunneling rate is more mobile and will move first, indeed the heavier atom
tunnels first to the left well. During the fast inclination also states with two
atoms at the same lattice site are occupied, which is reflected by a reduction
of the mean distance $\sqrt{\mean{\rho_x^2}}$. The occupation probability of
states above the first Bloch band is high (see top of Fig.~\ref{fig:fastincline}), 
and thus the behavior cannot be
described within a single-band approximation of the Hubbard model.

%%%%%%%%%%%%%%%%%%%%%%%%%%%%%%%%%%%%%%%%%%%%%%%%%%
 \subsection{Harmonic perturbation}
%%%%%%%%%%%%%%%%%%%%%%%%%%%%%%%%%%%%%%%%%%%%%%%%%%

\begin{figure}[htp]
 \centering
(a)\hfill\phantom{x}
 \includegraphics[width=\imgwidth]{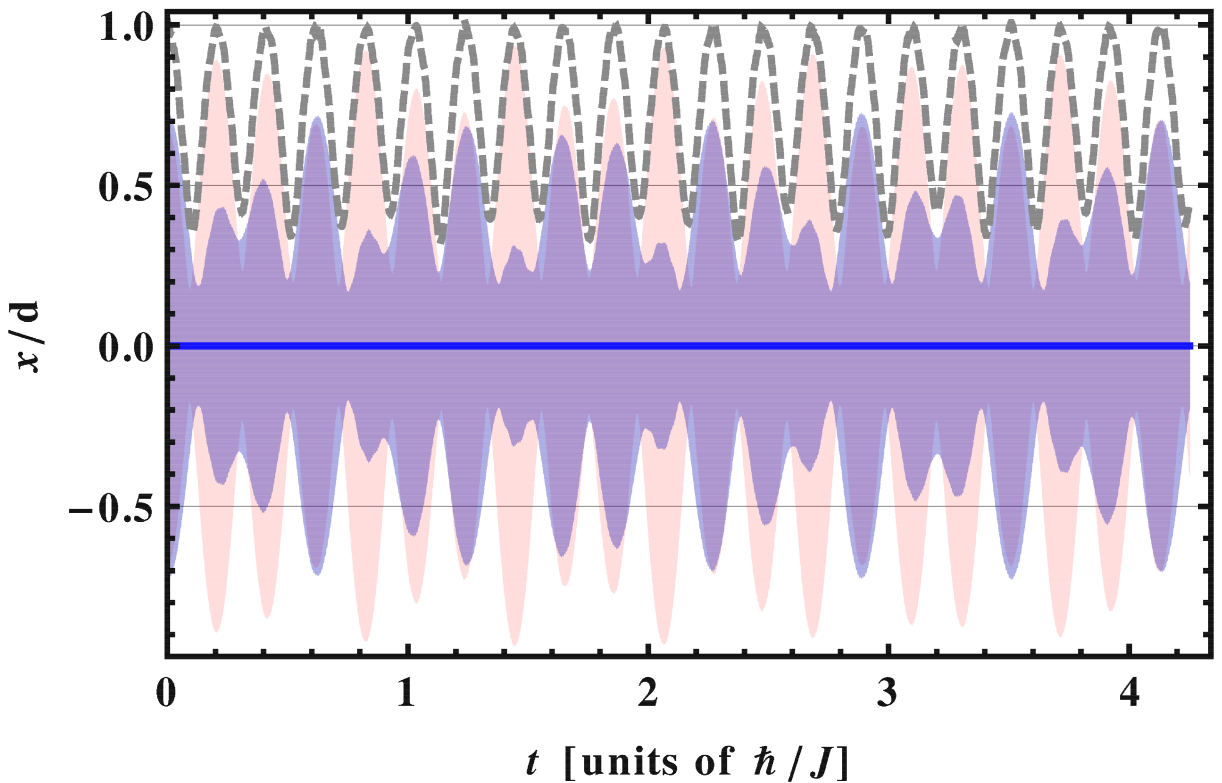}
(b)\hfill\phantom{x}
 \includegraphics[width=\imgwidth]{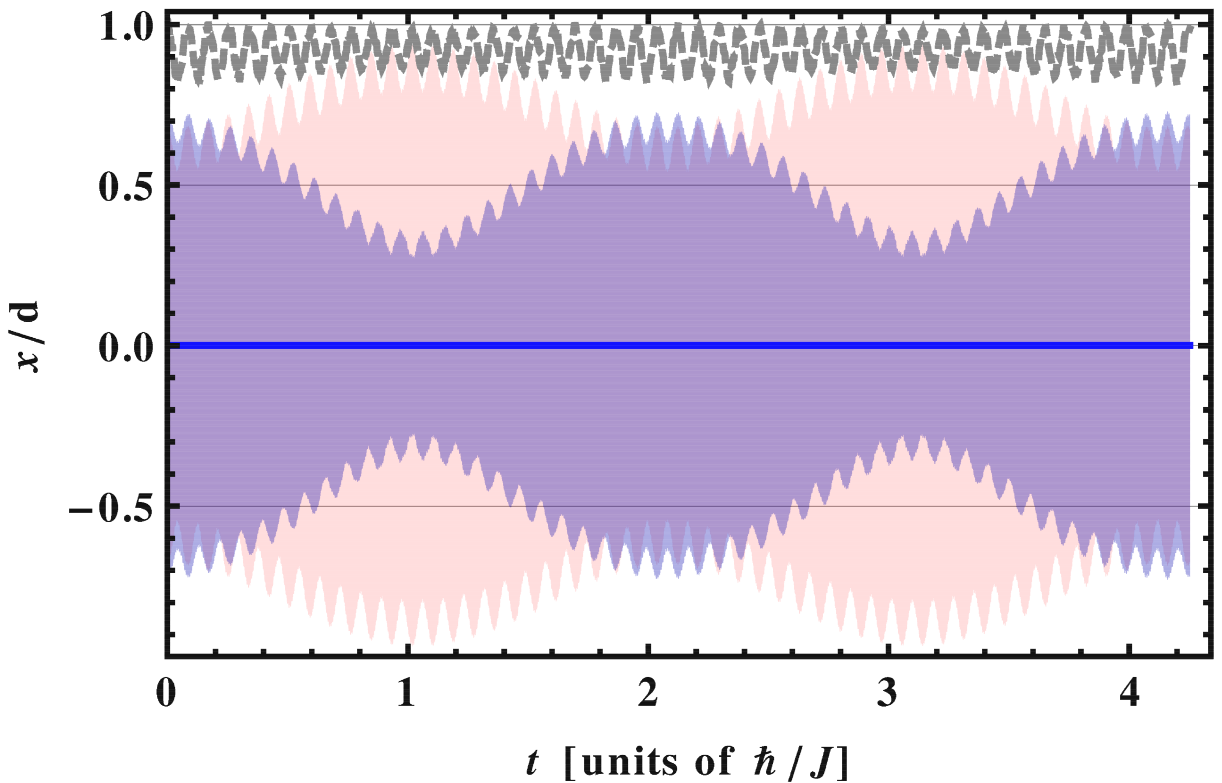}
 \caption{(color online) Time-dependent behavior for the sudden turn-on of an
additional harmonic confinement.
 (a) For $\hat W = 82\, J_1(\hat x_1^2 + \hat x_2^2)/d^2= 0.01\, \hbar\omega_1 (\hat x_1^2 +
\hat x_2^2)/d^2$ oscillations between bound and unbound states appear. 
 (b) For stronger confinement $\hat W = 164\, J_1 (\hat x_1^2 + \hat x_2^2)/d^2= 0.02\,
\hbar\omega_1 (\hat x_1^2 + \hat x_2^2)/d^2$ the bound-state occupation is much weaker,
however the particles tunnel alternating between the central and outer wells.
 Legend as in Fig.~\ref{fig:incline}.
  }
 \label{fig:bend}
\end{figure}

In experiments optical lattices are not infinite but the atoms are normally confined by
an additional weak harmonic potential. In the following the effect of the sudden
activation of such a harmonic potential $\hat W = A (\hat x_1^2 + \hat x_2^2)/d^2$ is
studied. This perturbation does not break the symmetry of the potential and the mean
position of the atoms remains at $x/d = 0$. However, as one can see in Fig.~\ref{fig:bend} (a) 
for a certain strength of the harmonic perturbation the system oscillates between unbound states ($\sqrt{\mean{\rho_x^2}}\approx d$)
and repulsively bound states ($\sqrt{\mean{\rho_x^2}}\approx 0.5 d$) \cite{cold:wink06} that are in resonance.
These oscillations are also visible in the uncertainty of the atoms' positions. 
For an increased harmonic perturbation 
no repulsively bound state is in resonance with the unbound state. 
Hence, as shown in Fig.~\ref{fig:bend}(b) the atoms oscillate predominantly between delocalized
states and states localized at the central lattice site. 
Since the atoms repel each other, the oscillations are exactly opposing each other.
The off-resonant coupling to the bound state leads
to small and fast oscillations of the mean distance $\sqrt{\mean{\rho_x^2}}$ between $0.8 d$ and $1.0 d$.

%%%%%%%%%%%%%%%%%%%%%%%%%%%%%%%%%%%%%%%%%%%%%%%%%%
 \section{Conclusion and outlook}
 \label{sec:conclusion}
 A theoretical approach for the full non-perturbative time-dependent description of two interacting
particles in an optical lattice was introduced. A comparison with analytical
results shows the possibility to perform high-precision analyses. Example
calculations for \LiLi in a three-well optical lattice where performed, 
demonstrating the possibility to analyze this complex six-dimensional system in 
terms of several expectation values. 
It was shown how the atoms are separated by a slowly increasing acceleration of the system and
how the system reacts upon suddenly stopping the acceleration.
It was also demonstrated that a fast acceleration of the lattice leads to a strong occupation
of states above the first Bloch band, which marks the break down of the usually adopted single-band Hubbard models.
As finally shown, a weak harmonic perturbation can have an important impact if the system encounters a resonance 
between bound and unbound states.

The use of a spectral method, i.\,e.\ expanding the time-dependent wavefunction in a basis
of eigenfunctions of some underlying Hamiltonian, offers a large degree of flexibility.
For example, by modifying the underlying Hamiltonian the here-presented system of two neutral atoms can be
easily generalized to other particles, such as ions, or dipoles. Also the
external potential is flexible enough to describe a large class of systems like
quantum dots or one- and two-dimensional optical traps. In the future we intend
to analyse and develop with the presented procedure schemes for the fast and
high-fidelity manipulation of small quantum systems.
%%%%%%%%%%%%%%%%%%%%%%%%%%%%%%%%%%%%%%%%%%%%%%%%%%

%\bibliographystyle{plainnat}
%\bibliography{cold}
%\bibliographystyle{/usr/share/texmf/tex/latex/revtex41/bibtex/bst/revtex/apsrev4-1}
%\bibliography{/data/amo/group/tex/bib/journals,/data/amo/group/tex/bib/cold}
%
%merlin.mbs apsrev4-1.bst 2010-07-25 4.21a (PWD, AO, DPC) hacked
%Control: key (0)
%Control: author (72) initials jnrlst
%Control: editor formatted (1) identically to author
%Control: production of article title (-1) disabled
%Control: page (0) single
%Control: year (1) truncated
%Control: production of eprint (0) enabled
%
 
\end{document}